\documentclass[aps,pra,twocolumn,amsmath,amssymb,superscriptaddress,longbibliography]{revtex4-2}

\usepackage[english]{babel}
\usepackage{xcolor}
\usepackage{amssymb} 
\usepackage{dcolumn}
\usepackage{bm}
\usepackage{graphicx}
\usepackage{amsmath}
\usepackage{braket}
\usepackage{graphicx}        

\graphicspath{{pict/}{}}

\usepackage[normalem]{ulem}
\usepackage{dcolumn}
\usepackage{bm}
\usepackage{balance}
\usepackage{mathrsfs}
\usepackage[pdfstartview=FitH, CJKbookmarks=true, bookmarksnumbered=true, bookmarksopen=true, colorlinks=true, pdfborder=001, citecolor=blue, linkcolor=blue, urlcolor=blue, linktocpage=true] {hyperref}

\makeatletter

\addto\captionsenglish{
	
}

\begin{document}

\preprint{APS/123-QED}


\title{Chirality-polarized superradiance induced by quantized orbital-angular-momentum transfer in a ring cavity}

\author{Yingqi Liu}
\affiliation{Guangdong Provincial Key Laboratory of Quantum Metrology and Sensing $\&$ School of Physics and Astronomy, Sun Yat-Sen University, Zhuhai 519082, China}

\author{Jingjun You}
\affiliation{Guangdong Provincial Key Laboratory of Quantum Metrology and Sensing $\&$ School of Physics and Astronomy, Sun Yat-Sen University, Zhuhai 519082, China}

	
\author{Jing Tang}
\email{jingtang@gdut.edu.cn}
\affiliation{School of Physics and Optoelectronic Engineering, Guangdong University of Technology, Guangzhou 510006, China}
\affiliation{Guangdong Provincial Key Laboratory of Sensing Physics and System Integration Applications, Guangdong University of Technology, Guangzhou, 510006, China}

\author{Yuangang Deng}
\email{dengyg3@mail.sysu.edu.cn}
\affiliation{Guangdong Provincial Key Laboratory of Quantum Metrology and Sensing $\&$ School of Physics and Astronomy, Sun Yat-Sen University, Zhuhai 519082, China}

\date{\today}

\begin{abstract}
We investigate a spin-$1/2$ Bose-Einstein condensate coupled to two counterpropagating modes of an optical ring cavity and driven by Laguerre-Gaussian Raman beams carrying quantized orbital angular momentum (OAM). We demonstrate that competition between OAM-selective Raman scattering channels controls collective superradiance through self-consistent atom--cavity feedback, linking atomic orbital motion to cavity-chirality selection. For unequal OAM transfer, one cavity mode dominates the superradiant response, with its chirality switchable by the Raman pump strengths. In contrast, equal OAM transfer allows both counterpropagating modes to coexist. The resulting chirality-polarized phases support distinct vortex and vortex--antivortex configurations with reconstructed synthetic gauge-field textures. Remarkably, contact interactions can reshape the vortex topology without changing the dominant cavity chirality, revealing distinct responses of atomic and photonic degrees of freedom. The predicted mode switching and vortex ordering can be probed through cavity output and spin-resolved imaging in a proposed $^{87}$Rb implementation. Our results establish quantized OAM transfer as a mechanism for controlling collective cavity emission beyond conventional atomic SOAMC.
\end{abstract}

\maketitle

\section{Introduction}\label{one}
Synthetic gauge fields generated by atom-light interactions provide a powerful platform for engineering topological and strongly correlated quantum matter in ultracold atomic gases~\cite{RevModPhys.72.1,RevModPhys.69.213,RevModPhys.52.175,RevModPhys.65.733,RevModPhys.94.045003,RevModPhys.73.663,RevModPhys.83.1523,Goldman_2014,RevModPhys.89.011004}. Coherent optical coupling between atomic internal states and center-of-mass motion enables the realization of effective gauge potentials in electrically neutral gases~\cite{PhysRevLett.102.130401,Lin2009,PhysRevLett.108.225304,Phuc2015}. Among various implementations, Raman-induced spin-orbit coupling (SOC) provides a versatile means of coupling atomic spin and momentum, giving rise to unconventional superfluid phases, nontrivial spin textures, and quantum phase transitions~\cite{Lin2011,PhysRevLett.109.095301,PhysRevLett.109.095302,PhysRevLett.109.115301,Fu2014,Lewenstein01032007,PhysRevLett.107.150403,PhysRevLett.108.225301,PhysRevLett.109.105303,PhysRevLett.110.235302,Zhai_2015,PhysRevLett.124.173203,PhysRevLett.125.195303}.

A particularly intriguing extension of conventional SOC is spin-orbital-angular-momentum coupling (SOAMC), realized by coupling atomic spin states to Laguerre-Gaussian (LG) Raman beams carrying quantized orbital angular momentum (OAM)~\cite{doi:10.1126/science.aao5392,PhysRevX.2.041011,RevModPhys.89.035004,Clark_2012,PhysRevLett.108.044801,PhysRevLett.121.113204,PhysRevA.91.063627,PhysRevLett.122.110402}. 
Unlike linear SOC, SOAMC preserves rotational symmetry while coupling the atomic spin to quantized orbital motion through discrete OAM transfer. The resulting angular-momentum-resolved Raman processes naturally obey OAM conservation and give rise to a wealth of phenomena, including quantized vortices, synthetic gauge fields, and topological spin textures in Bose-Einstein condensate (BEC) and degenerate Fermi gases~\cite{PhysRevLett.121.113204,PhysRevA.91.063627,PhysRevLett.122.110402,PhysRevA.91.033630,PhysRevA.93.013629,PhysRevA.102.063328,PhysRevA.105.063308,PhysRevA.105.023320,PhysRevResearch.2.033152,PhysRevResearch.6.033200,PhysRevA.92.033615,PhysRevLett.125.260407,PhysRevLett.126.193401}. Beyond these single-particle and equilibrium phenomena, the quantized OAM-transfer channels inherent to SOAMC provide a fundamentally new avenue for manipulating collective light-matter interactions through angular-momentum-selective Raman scattering, suggesting the possibility of engineering chirality-selective many-body phases in cavity quantum electrodynamics~\cite{PhysRevA.94.033627,PhysRevA.108.043310,PhysRevLett.121.250401,PhysRevLett.104.030401,PhysRevLett.108.035301,PhysRevLett.90.140403,Ray2014,doi:10.1126/science.1258289}.

Meanwhile, ultracold atoms coupled to optical cavities provide a versatile platform for studying collective light-matter interactions and nonequilibrium many-body physics~\cite{RevModPhys.85.553}. Collective coupling of an atomic ensemble to a common cavity mediates long-range interactions~\cite{doi:10.1126/science.aar3102,RevModPhys.95.035002}, leading to a broad range of emergent phenomena, including Dicke superradiance, self-organization, and supersolid phases~\cite{Brennecke2007,PhysRevResearch.7.013170,RevModPhys.95.035002,PhysRev.93.99,leonard2017monitoring,PhysRevLett.112.143007}. 
Moreover, the dissipative nature of the cavity field enables nondestructive monitoring of collective quantum dynamics~\cite{PhysRevLett.128.103201}. A paradigmatic example is the Dicke model, where an ensemble of two-level atoms collectively couples to a single electromagnetic mode and exhibits a superradiant phase transition~\cite{PhysRev.93.99,leonard2017monitoring,PhysRevLett.112.143007}. Experimentally, superradiant Rayleigh scattering from Bose-Einstein condensates and cavity-enhanced collective atomic recoil have been observed, accompanied by the redistribution of atoms among linear-momentum states~\cite{Inouye_1999,Slama_2007,Bux_2011}. In contrast, the present work considers stationary superradiant phases arising from cavity-mediated spin-orbital-angular-momentum coupling, where quantized OAM transfer links the condensate vortex structure to the selection of counterpropagating cavity modes.

 More recently, optical ring cavities supporting two counterpropagating running-wave modes have attracted considerable attention because they possess an additional cavity-chirality degree of freedom, enabling cavity-chirality-dependent ordering, emergent continuous symmetries, and supersolid phases, and other unconventional collective phenomena~\cite{PhysRevLett.120.123601,PhysRevLett.112.115302,PhysRevLett.122.190801,PhysRevLett.124.033601,PhysRevLett.131.173401,PhysRevResearch.5.013002}. The coexistence of two degenerate running-wave modes provides a unique setting for exploring chirality-dependent collective light-matter interactions and nonequilibrium many-body dynamics, thereby offering new opportunities for engineering directional collective emission and topological light-matter states~\cite{PhysRevLett.131.113602,qr8d-dh7g,lyu2025}.

Although SOAMC-induced vortex phases and ring-cavity self-organization have been investigated separately, how distinct quantized OAM-transfer channels affect the self-consistent occupation of counterpropagating cavity modes remains an open question. In particular, coupling different atomic angular-momentum sectors to separate cavity modes introduces a competition between the Raman scattering channels, allowing the condensate orbital configuration and cavity chirality to influence each other. This raises the possibility of controlling collective superradiant emission through quantized OAM transfer.

In this work, we investigate a spin-1/2 Bose-Einstein condensate coupled to a two-mode optical ring cavity and driven by LG Raman fields carrying distinct OAM. We show that angular-momentum-selective Raman scattering provides a mechanism for controlling the relative occupation of counterpropagating superradiant cavity modes. When the two Raman channels transfer unequal quantized OAM, their self-consistent coupling to the condensate favors one cavity chirality and allows switching between the two modes as the pump strengths are varied. By contrast, equal OAM transfer leads to the coexistence of both cavity modes in the parameter regime considered. The cavity-chirality selection is accompanied by distinct vortex and vortex--antivortex configurations and synthetic gauge-field textures, reflecting the coupling between atomic orbital motion and collective cavity emission. We further show that contact interactions can substantially modify the vortex topology without necessarily changing the dominant cavity chirality. These results establish a connection between quantized atomic OAM transfer and the mode-resolved superradiant response of a ring cavity.

This article is organized as follows. In Sec.~\ref{two}, we introduce the physical setup and derive the effective Hamiltonian describing cavity-mediated SOAMC in a spin-$1/2$ Bose-Einstein condensate. In Sec.~\ref{three}, we investigate the case of Laguerre-Gaussian Raman beams carrying distinct orbital angular momenta and characterize the resulting superradiant phases through the ground-state phase diagram, cavity and condensate properties, and the associated vortex structures and synthetic gauge fields. Section~\ref{four} is devoted to the effects of two-body interactions, where we examine the interaction-driven evolution and stability of the superradiant phases together with their topological properties. In Sec.~\ref{five}, we consider the equal-orbital-angular-momentum configuration and analyze the restoration of chirality-unpolarized superradiance and the corresponding Raman scattering mechanism. Finally, Sec.~\ref{six} summarizes our main findings and discusses future perspectives.

\begin{figure}[t]
\includegraphics[width=1.0\linewidth]{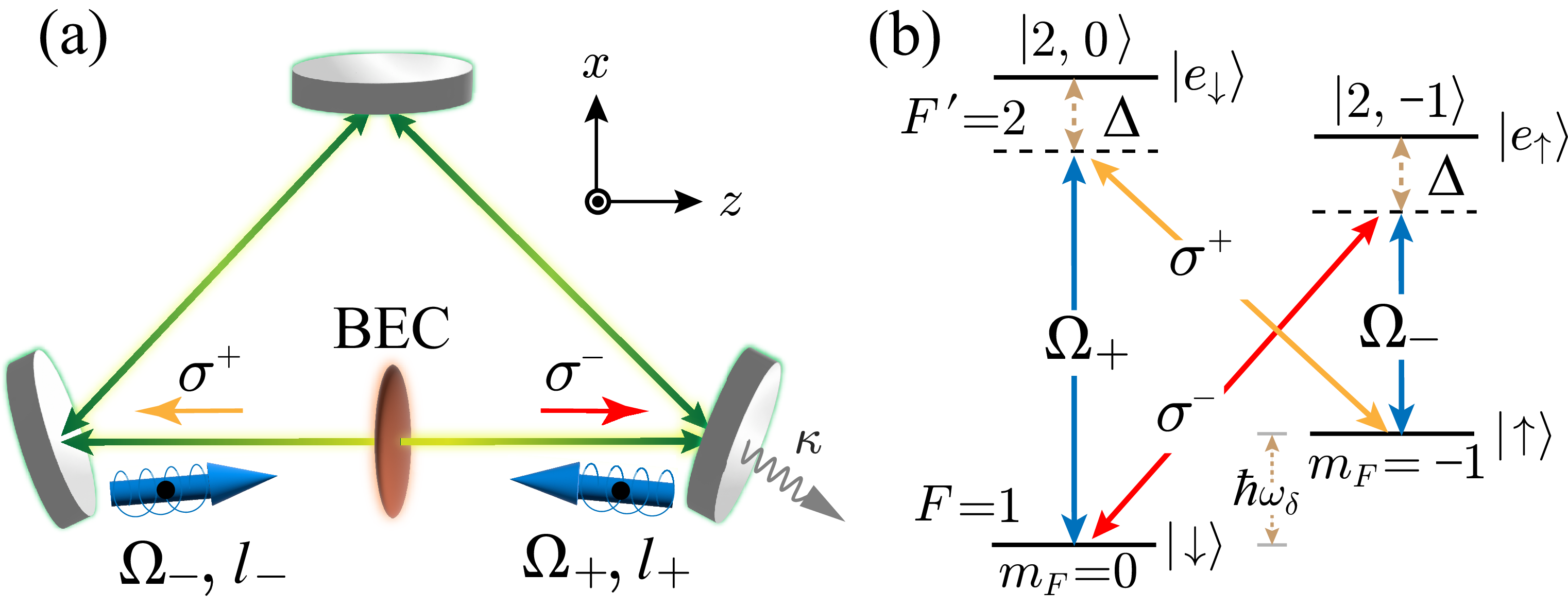}
\caption{(a) Schematic of the cavity-mediated SOAMC, with external LG Raman beams nearly copropagating with the corresponding cavity modes at a small crossing angle $\theta$. (b) Relevant Raman transitions between the $5S_{1/2}(F=1)$ and $5P_{3/2}(F'=2)$ manifolds of $^{87}$Rb. Two hyperfine ground states are coupled to two excited states through cavity-assisted Raman processes. The cavity fields provide the $\sigma^\pm$ polarization components, while two $\pi$-polarized LG Raman beams carrying orbital angular momenta $\hbar l_-$ and $\hbar l_+$ drive the transitions $|\uparrow\rangle\leftrightarrow|e_\uparrow\rangle$ and $|\downarrow\rangle\leftrightarrow|e_\downarrow\rangle$, respectively. The resulting Raman processes couple atomic orbital motion to the counterpropagating cavity modes.}
\label{Fig1}
\end{figure}

\section{Model and Hamiltonian}\label{two}

We consider an $N$-atom quasi-two-dimensional $^{87}$Rb Bose-Einstein condensate confined in the $xy$ plane and coupled to a two-mode optical ring cavity, as illustrated in Fig.~\ref{Fig1}. The condensate is prepared in the $5S_{1/2}(F=1)$ hyperfine ground-state manifold, where two Zeeman sublevels $\ket{\uparrow}=\ket{F=1,m_F=-1}$ and $\ket{\downarrow}=\ket{F=1,m_F=0}$ constitute an effective spin-$1/2$ system. Here $F$ and $m_F$ denote the total atomic angular momentum and its projection along the quantization axis, respectively. A bias magnetic field ${\bf B}$ applied along the $y$ direction defines the quantization axis and generates a Zeeman splitting $\hbar\omega_Z$ between the two spin states.  For sufficiently large ${\bf B}$, the remaining state $\ket{F=1,m_F=1}$ is shifted far from the spin-$1/2$ manifold by the quadratic Zeeman effect and can therefore be neglected in the low-energy dynamics~\cite{Lin2011}. 

The optical ring cavity supports two counterpropagating running-wave modes propagating along the $\mp z$ directions, corresponding to clockwise (CW) and counterclockwise (CCW) cavity fields. Both cavity modes are taken to be linearly polarized along the $x$ direction. With the quantization axis defined by the bias magnetic field along $y$, each cavity mode contains equal-magnitude $\sigma^+$ and $\sigma^-$ polarization components in the $y$-quantization basis, allowing transitions with $\Delta m_F=\pm1$. The $\sigma^\pm$ labels therefore refer to the atomic selection rules rather than distinct circular polarizations of the counterpropagating cavity modes. These cavity modes couple the two ground spin states to the electronically excited states $|e_\sigma\rangle$ ($\sigma=\uparrow,\downarrow$) in the $5P_{3/2}(F'=2)$ manifold. The relevant excited states are $|e_\uparrow\rangle=|F'=2,m_F'=-1\rangle$ and $|e_\downarrow\rangle=|F'=2,m_F'=0\rangle$.  The cavity fields couple to the transitions $|\uparrow\rangle\leftrightarrow|e_\downarrow\rangle$ and $|\downarrow\rangle\leftrightarrow|e_\uparrow\rangle$ through their $\sigma^+$ ($\Delta m_F=+1$) and $\sigma^-$ ($\Delta m_F=-1$) components, respectively, defined relative to the $y$-quantization axis. The $y$-polarized LG pumps drive the complementary $\pi$ transitions with $\Delta m_F=0$. The corresponding position-dependent coupling amplitudes are modeled as $g_\pm(\mathbf r)=g_0e^{\pm ik_Lz}$, where $k_L=2\pi/\lambda$ is the cavity wave vector and $\lambda$ is the optical wavelength. The subscripts $\pm$ distinguish the two counterpropagating cavity modes.

To generate cavity-mediated SOAMC, two classical Laguerre-Gaussian Raman fields couple the complementary atomic transitions $|\uparrow\rangle\!\leftrightarrow\!|e_{\uparrow}\rangle$ and $|\downarrow\rangle\!\leftrightarrow\!|e_{\downarrow}\rangle$. These fields are externally applied pumps rather than cavity eigenmodes and overlap with the counterpropagating cavity fields at the atomic cloud. The two LG beams propagate nearly along the $\pm z$ directions, making a small angle $\theta$ with their respective copropagating cavity modes, and carry quantized orbital angular momenta $\hbar l_{\pm}$. In the near-copropagating approximation, their Rabi frequencies are taken as~\cite{PhysRevA.91.063627}
\begin{equation}
    \Omega_{\pm}({\bf r})
    = \Omega_{\pm}
    \left(\frac{r}{w_0}\right)^{|l_{\pm}|}
    e^{-r^2/w_0^2}
    e^{i l_{\pm}\varphi}
    e^{\pm i k_L z},
    \label{OmegaLG}
\end{equation}
where ${\bf r}=(r,\varphi,z)$ denotes the cylindrical coordinate and $w_0$ is the beam waist. The amplitudes $\Omega_{\xi}$ are proportional to the pump strengths, while the azimuthal phase factors $e^{il_{\pm}\varphi}$ impart quantized OAM to the atoms during the Raman process.

The external LG Raman pumps nearly copropagate with their respective cavity modes at a small crossing angle $\theta$, allowing spatial overlap at the condensate without requiring the pumps to be cavity eigenmodes~\cite{Ostermann_2019}. The residual wave-vector mismatch $\Delta\mathbf{k}_\pm=\mathbf{k}_{p,\pm}-\mathbf{k}_{c,\pm}$ is assumed to satisfy $|\Delta\mathbf{k}_{\pm,\perp}|R\ll1$, where $R$ denotes the transverse condensate radius. Under these conditions, the pump--cavity spatial phases approximately cancel, leaving the azimuthal factor $e^{il_\pm\phi}$ responsible for SOAMC. The independently tunable pump frequencies $\omega_{p,\pm}$ satisfy the corresponding two-photon Raman resonance conditions without requiring resonance of the pumps with the cavity modes. This configuration requires independent optical access and careful mode matching. For simplicity, equal effective cavity detunings and decay rates are assumed throughout.

The above configuration is modeled by two selected cavity-assisted Raman scattering channels connecting $\ket{\uparrow}$ and $\ket{\downarrow}$, assuming that other polarization-allowed Raman pathways are sufficiently off-resonant to be neglected. For a large one-photon detuning $\Delta$ satisfying $|\Delta|\gg |g_\pm|,|\Omega_\pm|$, the electronically excited states $|e_\sigma\rangle$ can be adiabatically eliminated. Spontaneous scattering is neglected within this approximation, provided that its residual rate remains small compared with the relevant atomic dynamical rates. The resulting two-photon Raman processes generate an effective spin-flip coupling accompanied by quantized orbital-angular-momentum transfer from the LG beams to the condensate. Consequently, the cavity field, the atomic spin, and the atomic orbital motion become dynamically coupled, thereby realizing cavity-mediated SOAMC.

Although the two counterpropagating cavity modes may have different resonance frequencies, we consider equal effective cavity detunings and decay rates in the following calculations for simplicity. Including short-range contact interactions, the many-body Hamiltonian describing the cavity-mediated SOAMC condensate is given by
\begin{align}
\hat{\cal H}=&\hbar \Delta_{a}(\hat{a}_-^{\dagger}\hat{a}_- + \hat{a}_+^{\dagger}\hat{a}_+)\nonumber \\
			&+{\sum_{\sigma\sigma^{\prime}}\int{d\mathbf{r}{\hat{\psi}}_{\sigma}^{\dag}(\mathbf{r})[\hat{h}_{\sigma\sigma^{\prime}}+V(\mathbf{r})\delta_{\sigma\sigma^{\prime}}]{\hat{\psi}}_{\sigma^{\prime}}(\mathbf{r})}}\nonumber \\
			&+{\sum_{\sigma\sigma^{\prime}}g_{\sigma\sigma^{\prime}}\int{d\mathbf{r}{\hat{\psi}}_{\sigma}^{\dag}(\mathbf{r}){\hat{\psi}}_{\sigma^{\prime}}^{\dag}(\mathbf{r}){\hat{\psi}}_{\sigma^{\prime}}(\mathbf{r}){\hat{\psi}}_{\sigma}(\mathbf{r})}},
	\label{eq1}
\end{align}
where $\hat\psi_{\sigma}$ is the bosonic field operator for spin-$\sigma$ atoms, and $\hat a_\pm$ annihilates photons in the two counterpropagating cavity modes. Here $\Delta_a$ denotes the cavity-pump detuning, $V(\mathbf r)$ is the external trapping potential, and $g_{\sigma\sigma'}=4\pi\hbar^2a_{\sigma\sigma'}/M$ characterizes the contact interaction between atoms in spin states $\sigma$ and $\sigma'$, with $a_{\sigma\sigma'}$ the corresponding $s$-wave scattering lengths and $M$ the atomic mass. Explicitly,  the effective single-particle Hamiltonian reads 
\begin{eqnarray}
\hat {\boldsymbol h} =\frac{{\mathbf p}^{2}}{2M}\hat I+
\hbar\left(\!
\begin{array}{cc}
\hat{U}_- +{M}_+  +  {\delta}/2 & \eta_+ \hat{a}_+ +\eta_- \hat{a}_-^\dag   \\
\eta_+^* \hat{a}^\dag_+ +\eta_-^* \hat{a}_- & \hat{U}_+ +{M}_--{\delta}/2
\end{array} 
\!\right)\!,\;\label{singleh}
\end{eqnarray}
where $\hat{U}_{\pm}=U_0 \hat{a}_{\pm}^{\dagger} \hat{a}_{\pm} $ with $U_0=-\mathrm{g}_0^2/\Delta$ being the optical Stark shift of the cavity field, ${M}_{\pm}({\bf r})=-|\Omega_{\pm}({\bf r})|^2 /\Delta$ is the Stark shift of classical LG pump field, $\delta$ is the tunable two-photon detuning, and $\eta_{\pm}=-{g^*_{\pm} {\Omega}_{\pm}({\bf r})}/{\Delta}$ are the maximum scattering rate for Raman processes in cavity-coupled BECs. Unlike cavity-mediated dynamical spin-orbit coupling~\cite{PhysRevLett.112.143007,PhysRevA.89.011602,you2025spin,PhysRevLett.123.160404,PhysRevResearch.5.013002}, the present scheme employs two Raman channels carrying different quantized orbital angular momenta. These channels couple distinct atomic orbital states to the counterpropagating cavity modes, linking the condensate vortex structure to cavity chirality. Their competition provides the microscopic mechanism for chirality-selective superradiance.

For a sufficiently large cavity-pump detuning $|\Delta_a|\gg|\eta_\pm|$, we assume that the cavity-field relaxation is much faster than the relevant atomic dynamics, allowing the cavity fields to follow the condensate motion adiabatically~\cite{RevModPhys.85.553,mivehvar2021cavity}. The cavity fields  $\hat a$ in $\hat{\cal H}$ can therefore be adiabatically eliminated through the dynamical equations for the field operators, 
\begin{align}
i \dot{\hat{a}}_+ =& (\Delta_{a} +U_{0} N -i\kappa)\hat{a}_+ +   \int dr \: \: \eta_{+}^* \hat{\psi}^{\dagger}_\uparrow \hat{\psi}_\downarrow, \nonumber \\
i \dot{\hat{a}}_- =& (\Delta_{a} +U_{0} N -i\kappa)\hat{a}_- +  \int dr \: \: \eta_{-} \hat{\psi}^{\dagger}_\downarrow \hat{\psi}_\uparrow, 
\end{align}
The cavity-field relaxation time is characterized by $\tau_c=\kappa^{-1}$. The adiabatic approximation requires $\tau_c\ll\tau_{\mathrm{COM}}$, where $\tau_{\mathrm{COM}}$ is the characteristic timescale of the condensate dynamics. Spontaneous scattering from the electronically excited states is neglected within the large-detuning approximation. The steady-state intracavity field amplitudes are therefore given by
\begin{align}
\hat{a}_+  &= -\cfrac{\int dr\: \eta_{+}^* \hat{\psi}^{\dagger}_ \uparrow\hat{\psi}_\downarrow}{\Delta_{a} +U_{0} N -i\kappa},  \nonumber \\
\hat{a}_- &=  -\cfrac{\int dr\: \eta_{-} \hat{\psi}^{\dagger}_\downarrow \hat{\psi}_\uparrow}{\Delta_{a} +U_{0} N -i\kappa},  \label{sscav}
\end{align}
Equations (\ref{sscav}) describe the deterministic cavity response, with cavity-field quantum noise neglected. The condition $|\Delta_a+U_0N|\gg\kappa$ characterizes the dispersive regime considered here.

Substituting the steady-state cavity fields into the Heisenberg equations of motion, $i\hbar\partial\hat\psi_\sigma(\mathbf r)/\partial t=[\hat\psi_\sigma(\mathbf r),\hat H]$, we adopt the mean-field approximations $\hat a_\pm\rightarrow\alpha_\pm=\langle\hat a_\pm\rangle$ and $\hat\psi_\sigma(\mathbf r)\rightarrow\psi_\sigma(\mathbf r)=\langle\hat\psi_\sigma(\mathbf r)\rangle$. The cavity amplitudes remain self-consistently coupled to the atomic spin coherences, retaining cavity-mediated feedback while neglecting quantum fluctuations and connected atom-cavity correlations beyond the mean-field level. This yields the coupled Gross-Pitaevskii equations,
\begin{equation}
	i\hbar \frac{\partial \psi_{\sigma}}{\partial t}
	=
	\sum_{\sigma'}
	\left(
	\hat h_{\sigma\sigma'}
	+
	V(\mathbf{r})\delta_{\sigma\sigma'}
	+
	\frac{4\pi\hbar^2 a_{\sigma\sigma'}}{M}
	\psi_{\sigma'}^{*}\psi_{\sigma}
	\right)
	\psi_{\sigma'},
	\label{eq:GPE}
\end{equation}
The expectation values $\alpha_{\pm}=\langle\hat a_{\pm}\rangle$ serve as the order parameters of the superradiant phase transition. A finite intracavity field ($\alpha_{\pm}\neq0$) signals the onset of the self-ordering superradiant phase, whereas $\alpha_{\pm}=0$ corresponds to the normal phase.

A key feature of the present setup is that the two cavity-assisted Raman pathways are associated with opposite cavity chiralities and distinct OAM-transfer channels. When $l_-\neq l_+$, the two Raman processes couple to different angular-momentum sectors of the condensate, allowing one cavity chirality to be selectively enhanced while suppressing the other. Consequently, angular-momentum-selective Raman scattering provides the microscopic mechanism underlying the chirality-polarized superradiance discussed below.

The ground-state properties are determined self-consistently together with the steady-state cavity fields obtained from Eq.~(\ref{sscav}). Throughout this work, we consider an experimentally accessible parameter regime for a $^{87}$Rb Bose-Einstein condensate.
The ground-state configurations are obtained by imaginary-time evolution of the coupled Gross-Pitaevskii equations [Eq.~(\ref{eq:GPE})], which is equivalent to minimizing the mean-field energy functional $\langle\hat{\mathcal H}\rangle$.\par

We consider a condensate of $N=10^4$ atoms initially prepared in the state $\ket{\uparrow}$ and confined in a circular optical box potential in the $xy$ plane with radius $R\simeq3.06\,\mu\mathrm{m}$~\cite{PhysRevLett.110.200406,Navon2016}. The transverse cavity-mode profile is approximated as uniform across the condensate, assuming a cavity waist $w_c\gg R$. Along the $z$ direction, the condensate is tightly confined by a harmonic potential with frequency $\omega_z=2\pi\times100\,\mathrm{kHz}$. Assuming that the axial motion is frozen to the harmonic-oscillator ground state, the condensate wave functions can be factorized as $\psi_\sigma(\mathbf r)=\phi_\sigma(x,y)\phi_z(z)$, where $\phi_z(z)=(\pi a_z^2)^{-1/4}\exp[-z^2/(2a_z^2)]$ and $a_z=\sqrt{\hbar/(M\omega_z)}\simeq34.1\,\mathrm{nm}$. Integrating out the axial coordinate yields an effective quasi-two-dimensional system with contact interaction strengths
\begin{equation}
g_{\sigma\sigma'}^{\mathrm{2D}}
=\frac{g_{\sigma\sigma'}}{\sqrt{2\pi}a_z}
=\frac{\sqrt{8\pi}\hbar^2a_{\sigma\sigma'}}{Ma_z},
\end{equation}
where $g_{\sigma\sigma'}=4\pi\hbar^2a_{\sigma\sigma'}/M$ and $a_{\sigma\sigma'}$ denotes the three-dimensional $s$-wave scattering length. The numerical simulations are performed by solving the resulting two-dimensional Gross-Pitaevskii equations, with the axial degree of freedom integrated out. The reduced condensate wave functions are normalized according to $\sum_\sigma\int d^2\boldsymbol{\rho}\,|\phi_\sigma(\boldsymbol{\rho})|^2=N$. The contact interactions are assumed to be spin independent, $a_{\uparrow\uparrow}=a_{\downarrow\downarrow}=a_{\uparrow\downarrow}=100.4a_B$, where $a_B$ is the Bohr radius, corresponding to an SU(2)-symmetric interaction regime. The single-photon recoil energy is $E_L=\hbar^2k_L^2/(2M)$, with $E_L/\hbar\simeq2\pi\times3.5\,\mathrm{kHz}$ for an optical wavelength of $\lambda=803.2\,\mathrm{nm}$. Unless otherwise specified, we choose $\hbar\kappa=50E_L$, $\hbar U_0=2E_L$, $\delta=0$, and $\Delta_a=-U_0N/2$.

Throughout this work, we focus on the unequal-OAM configuration with $l_-=1$ and $l_+=2$, for which the two cavity-assisted Raman processes transfer different quantized orbital angular momenta. As demonstrated below, this angular-momentum selectivity fundamentally modifies the collective superradiant response, giving rise to chirality-polarized superradiance and a rich family of topological superradiant phases.

\section{Results}\label{three}

\begin{figure}[t]
	\includegraphics[width=1.0\linewidth]{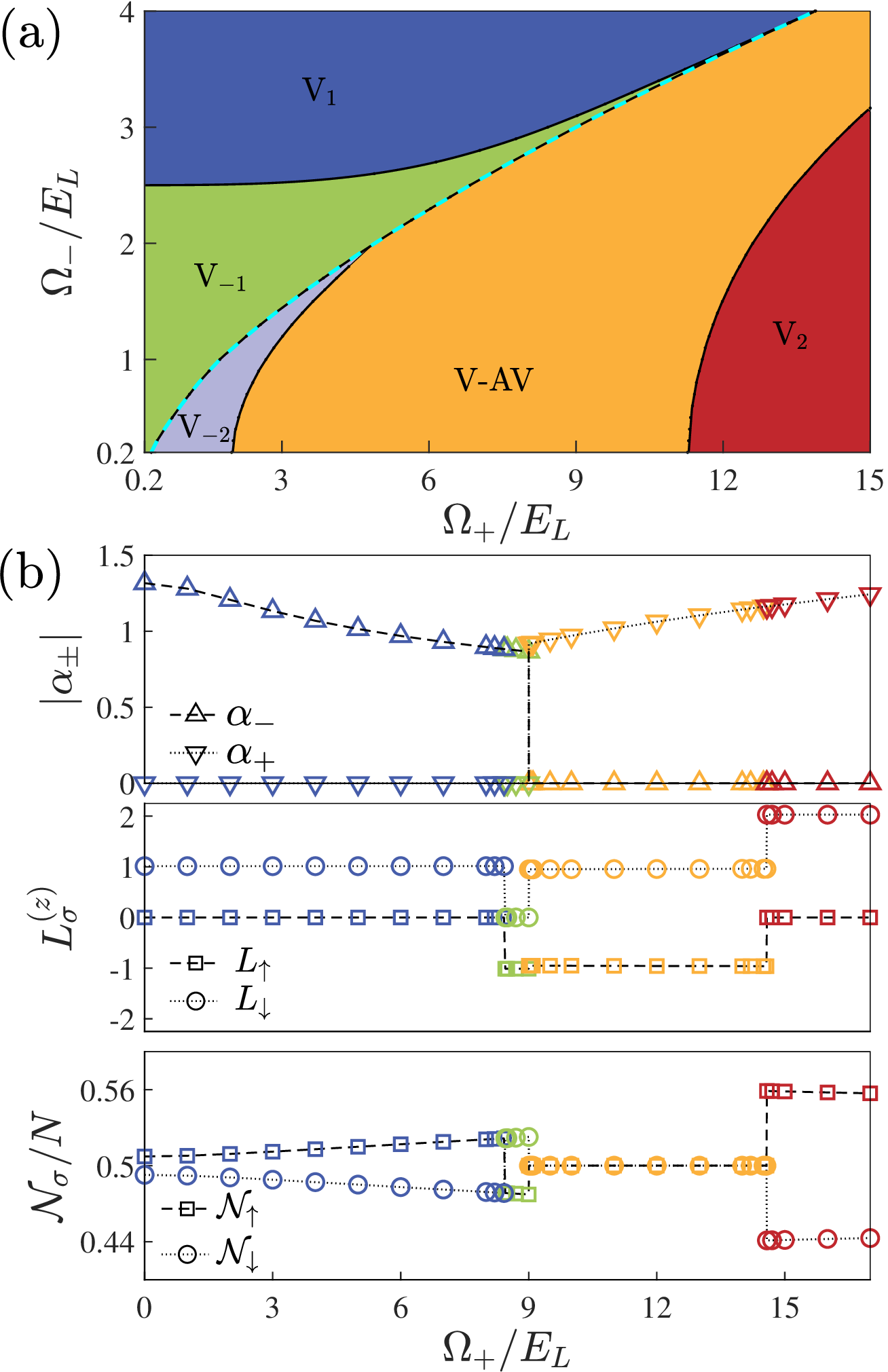}
	\caption{(Color online)(a) Ground-state phase diagram in the $\Omega_+$--$\Omega_-$ plane for unequal OAM-transfer channels ($l_-=1$, $l_+=2$). Five topological superradiant phases are found: the vortex phases $\mathrm{V}_{-2}$, $\mathrm{V}_{-1}$, $\mathrm{V}_{1}$, and $\mathrm{V}_{2}$, together with the vortex-antivortex phase $\mathrm{V}$--$\mathrm{AV}$. The cyan dashed curve marks the boundary between superradiant regions dominated by opposite cavity chiralities. (b) Steady-state cavity amplitudes $|\alpha_\pm|$ (upper panel), mean orbital angular momenta $L_\sigma^{(z)}$ (middle panel), and normalized spin populations $\mathcal{N}_\sigma/N$ (lower panel) as functions of $\Omega_+$ for $\Omega_-=3E_L$. Upward and downward triangles denote $|\alpha_-|$ and $|\alpha_+|$, respectively, while squares and circles represent the spin-up and spin-down components in the middle and lower panels. The colors correspond to the superradiant phases in (a). The evolution of the order parameters reveals the switching of the superradiant cavity chirality accompanied by successive vortex phase transitions.}
	\label{Fig2}
\end{figure}

Figure~\ref{Fig2}(a) presents the ground-state phase diagram in the $\Omega_+$-$\Omega_-$ plane for unequal OAM-transfer channels ($l_-=1$ and $l_+=2$). Above the superradiant threshold, a finite coherent intracavity field emerges together with vortex ordering in the condensate. These two features are linked by cavity-assisted Raman scattering, which transfers quantized orbital angular momentum to the atoms, while the resulting atomic spin coherences determine the intracavity field amplitudes self-consistently. Here, ``superradiance'' refers to cavity-induced collective self-organization, characterized by a nonzero intracavity order parameter and a corresponding reorganization of the condensate. The resulting vortex and vortex--antivortex configurations are stationary self-consistent solutions of the effective mean-field equations, rather than transient superradiant scattering dynamics.

A striking feature of the phase diagram is that superradiant emission is always dominated by a single running-wave cavity mode. Throughout the entire superradiant regime, the steady-state cavity amplitudes satisfy either $|\alpha_+/\alpha_-|\gg1$ or $|\alpha_-/\alpha_+|\gg1$, indicating that one cavity chirality is macroscopically occupied while the opposite chirality remains strongly suppressed. This self-consistent selection of a dominant cavity mode under unequal OAM transfer is the defining feature of chirality-polarized superradiance in our system.

The cyan dashed curve in Fig.~2(a) separates the two superradiant regions with opposite cavity chiralities. On the left side of the boundary, superradiant emission is dominated by the CCW cavity mode, whereas on the right side it switches to the CW mode. This boundary is consistent with the switching of the cavity-field amplitudes shown in Fig.~2(b). Correspondingly, five topological superradiant phases emerge, distinguished by the selected cavity chirality and the mean orbital angular momenta $(L_{\uparrow}^{(z)},L_{\downarrow}^{(z)})$: the vortex phases $\mathrm{V}_{-1}$ and $\mathrm{V}_{1}$, characterized by $(-1,0)$ and $(0,1)$ and supported by the $\hat a_-$ mode, and the phases $\mathrm{V}_{-2}$, $\mathrm{V}$-$\mathrm{AV}$, and $\mathrm{V}_{2}$, characterized by $(-2,0)$, $(-1,1)$, and $(0,2)$, respectively, which are supported by the CW cavity mode. Each superradiant phase is therefore uniquely associated with a cavity chirality and a corresponding vortex topology.

The origin of the chirality selection can be traced to the angular-momentum-selective Raman scattering introduced by unequal OAM transfer. Because the two LG Raman beams carry different quantized orbital angular momenta ($l_-\neq l_+$), the corresponding cavity-assisted Raman processes couple to distinct orbital-angular-momentum sectors of the condensate. Their competition therefore favors one scattering channel while suppressing the other, leading simultaneously to the selection of a unique cavity chirality and the associated vortex configuration. Unlike conventional ring-cavity superradiance, where both counterpropagating cavity modes generally coexist and their occupations evolve continuously with the pump configuration~\cite{PhysRevLett.120.123601,Colella_2019}, quantized OAM transfer lifts the symmetry between the two Raman channels and locks the superradiant emission to a single cavity chirality.

\textit{Chirality-polarized superradiance}.---To gain further insight into the phase diagram, we examine the evolution of the order parameters along the representative cut $\Omega_-=3E_L$ shown in Fig.~\ref{Fig2}(b). For weak $\Omega_+$, only CCW cavity mode is macroscopically occupied, whereas the counterpropagating mode remains strongly suppressed, demonstrating chirality-polarized superradiance in the $\hat a_-$ channel. As $\Omega_+$ increases, $|\alpha_-|$ decreases continuously and vanishes abruptly at the phase boundary, accompanied by the rapid buildup of the complementary CW cavity mode. This abrupt exchange of the intracavity field identifies a switching between opposite cavity chiralities, driven by the competition between Raman scattering channels transferring different quantized orbital angular momenta to the condensate.

The chirality switching is accompanied by successive topological phase transitions. For weak $\Omega_+$, the Raman process associated with the $l_-=1$ LG beam stabilizes the singly quantized vortex phases $\mathrm{V}_{1}$ and $\mathrm{V}_{-1}$. As $\Omega_+$ increases, the condensate evolves from $\mathrm{V}_{1}$ to $\mathrm{V}_{-1}$ while the CCW cavity field is gradually suppressed. Beyond the critical point, the competing Raman channel associated with the $l_+=2$ LG beam becomes dominant, driving the system into the $\mathrm{V}$-$\mathrm{AV}$ and $\mathrm{V}_{2}$ phases. These transitions are accompanied by quantized changes in the OAM and a redistribution of the spin populations, reflecting the transfer of OAM between the two spin components.

The switching of cavity chirality originates from angular-momentum-selective Raman scattering. Because the two cavity-assisted Raman processes transfer different quantized orbital angular momenta ($l_-\neq l_+$), they couple to distinct orbital-angular-momentum sectors of the condensate. Their competition therefore breaks the symmetry between the two counterpropagating cavity modes and selects a single superradiant scattering channel. Consequently, each topological superradiant phase is uniquely associated with one cavity chirality, establishing chirality-polarized superradiance as a direct consequence of quantized OAM transfer. The resulting chirality selection bears some resemblance to the nonreciprocal Dicke model~\cite{PhysRevLett.131.113602}, where competing photonic channels couple asymmetrically to matter. The underlying mechanisms, however, are fundamentally different. In the present system, chirality polarization emerges intrinsically from quantized orbital-angular-momentum transfer and the associated angular-momentum selection rules, without requiring engineered asymmetric light-matter couplings. This contrasts sharply with previous studies~\cite{Ostermann_2019}: quantized OAM transfer can serve as a new symmetry-breaking mechanism to achieve chirality selectivity in cavity QED.

\begin{figure}[t]
\includegraphics[width=0.99\linewidth]{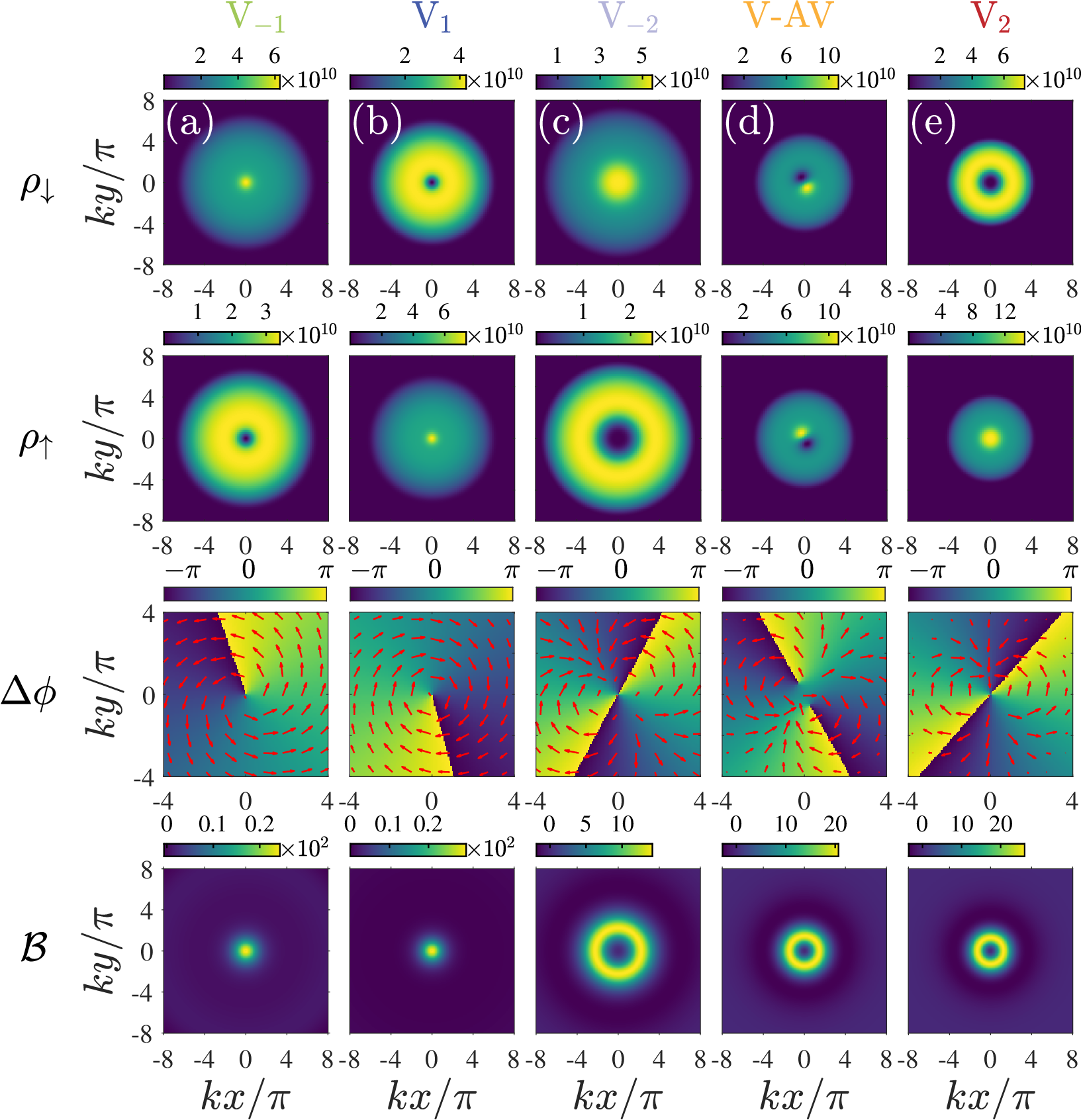}
\caption{
(Color online)
Representative ground-state properties of the five topological superradiant phases.
From top to bottom, the rows show the atomic column densities $\rho_{\downarrow}$ and $\rho_{\uparrow}$ (cm$^{-2}$), the relative phase $\Delta\phi=\phi_{\uparrow}-\phi_{\downarrow}$ (arrows denote the local spin texture), and the synthetic magnetic field $\bm{\mathcal B}$ (G).
The columns correspond to (a) $\mathrm{V}_{-1}$ with $(\Omega_-,\Omega_+)=(2,2)E_L$, (b) $\mathrm{V}_{1}$ with $(3,2)E_L$, (c) $\mathrm{V}_{-2}$ with $(1,2)E_L$, (d) $\mathrm{V}$-$\mathrm{AV}$ with $(1,8)E_L$, and (e) $\mathrm{V}_{2}$ with $(1,12)E_L$.
}
\label{Fig3}
\end{figure}

\textit{Density and gauge-field textures}.---To elucidate the physical origin of the distinct vortex topologies, we analyze the synthetic gauge field generated by cavity-mediated SOAMC. After subtracting the spin-independent terms, the spin-dependent part of the single-particle Hamiltonian can be written as
\begin{equation}
\frac{\hat h_s}{\hbar}
=
\frac{\delta_r}{2}\sigma_z
+
\left(
\Omega_Re^{i\phi_R}\sigma_+
+\mathrm{H.c.}
\right),
\label{eq4}
\end{equation}
where $\delta_r=\delta+U_-+M_+-U_+-M_-$ is the effective Raman detuning. The complex Raman coupling is defined by $\Omega_Re^{i\phi_R}=\eta_+\alpha_++\eta_-\alpha_-^*$, with $\alpha_\pm=\langle\hat a_\pm\rangle$ and $\Omega_R\geq0$. Thus, the Raman phase $\phi_R$ is determined by the coherent cavity fields and the azimuthal phases of the LG Raman beams.
The dressed-state eigenvalues of $\hat h_{\mathrm{s}}/\hbar$ are
\begin{equation}
\lambda_{\pm}
=
\pm\sqrt{\frac{\delta_r^2}{4}+\Omega_R^2}.
\label{eq:dressed-energy}
\end{equation}
We introduce the mixing angle $\beta$ through $\tan(2\beta)=2\Omega_R/\delta_r$, with $0\leq\beta\leq\pi/2$ and $\cos(2\beta)=\delta_r/\sqrt{\delta_r^2+4\Omega_R^2}$. A consistent choice of the normalized dressed states is
\begin{equation}
|\chi_-\rangle=
\begin{pmatrix}
-\sin\beta\\
\cos\beta\,e^{-i\phi_R}
\end{pmatrix},
\qquad
|\chi_+\rangle=
\begin{pmatrix}
\cos\beta\\
\sin\beta\,e^{-i\phi_R}
\end{pmatrix}
\label{eq:chi-minus}
\end{equation}
which satisfy $\hat h_{\mathrm{s}}|\chi_\pm\rangle=\hbar\lambda_\pm|\chi_\pm\rangle$.

Within the adiabatic dressed-state description, projection onto the lower branch gives the synthetic vector potential
\begin{equation}
\mathbf A
=
i\hbar\langle\chi_-|\nabla\chi_-\rangle
=
\hbar\cos^2\beta\,\nabla\phi_R.
\end{equation}
Away from singularities of the Raman phase, using $\nabla(\cos^2\beta)=-\sin(2\beta)\nabla\beta$, the corresponding synthetic magnetic field is
\begin{equation}
\bm{\mathcal B}
=
\nabla\times\mathbf A
=
-\hbar\sin(2\beta)
\left(
\frac{\partial\beta}{\partial x}
\frac{\partial\phi_R}{\partial y}
-
\frac{\partial\beta}{\partial y}
\frac{\partial\phi_R}{\partial x}
\right)\hat{\mathbf z}.
\label{eq5}
\end{equation}

Figure~\ref{Fig3} shows the density distributions, relative phases, and synthetic magnetic fields for the five representative superradiant phases. The synthetic gauge field undergoes a qualitative reconstruction across the different phases, establishing a direct correspondence between cavity chirality and vortex topology. The CCW-dominated phases $\mathrm{V}_{-1}$ and $\mathrm{V}_{1}$ are characterized by a synthetic magnetic field localized near the trap center. Consequently, the vortex core is pinned at the origin, while the relative phase exhibits a single $2\pi$ winding characteristic of a singly quantized vortex. Although both phases preserve continuous rotational symmetry, they possess opposite circulation and spin textures. As $\Omega_-$ increases, population is continuously transferred between the two spin components, driving the transition from the $\mathrm{V}_{-1}$ phase to the $\mathrm{V}_{1}$ phase. This evolution is accompanied by a gradual compression of the condensate resulting from the enhanced cavity-assisted Raman confinement.

A qualitatively different synthetic gauge-field texture emerges in the CW-dominated phases $\mathrm{V}_{-2}$, $\mathrm{V}$--$\mathrm{AV}$, and $\mathrm{V}_{2}$. In the vortex phases $\mathrm{V}_{-2}$ and $\mathrm{V}_{2}$, the synthetic magnetic field develops an annular profile with its maximum displaced from the trap center. This ring-shaped gauge field favors higher-orbital-angular-momentum states and stabilizes doubly quantized vortices while preserving rotational symmetry. The evolution from a centrally localized to an annular magnetic-field profile therefore signals the switching of the dominant cavity-assisted Raman channel and the associated quantized OAM transfer.

The intermediate $\mathrm{V}$-$\mathrm{AV}$ phase arises from the competition between the two cavity-assisted Raman channels. Near the phase boundary, the $l_-=1$ Raman channel remains sufficiently strong to sustain a singly quantized vortex, whereas the competing $l_+=2$ channel has already reconstructed the synthetic magnetic field into an annular profile but is not yet strong enough to stabilize a doubly quantized vortex. Consequently, the vortex is displaced from the trap center and becomes pinned near the maximum of the annular magnetic field, giving rise to the vortex-antivortex configuration and reducing the continuous rotational symmetry to a discrete one.

These density, phase, and gauge-field textures reveal a close connection between cavity chirality, synthetic gauge fields, and vortex topology. While distinct vortex configurations can occur within the same chirality-dominated regime, the competition between the two OAM-selective Raman channels leads to a reconstruction of the synthetic gauge-field texture. These results demonstrate how quantized OAM transfer jointly controls the collective cavity response and the orbital structure of the condensate.

 \begin{figure*}[t]
	\includegraphics[width=0.90\linewidth]{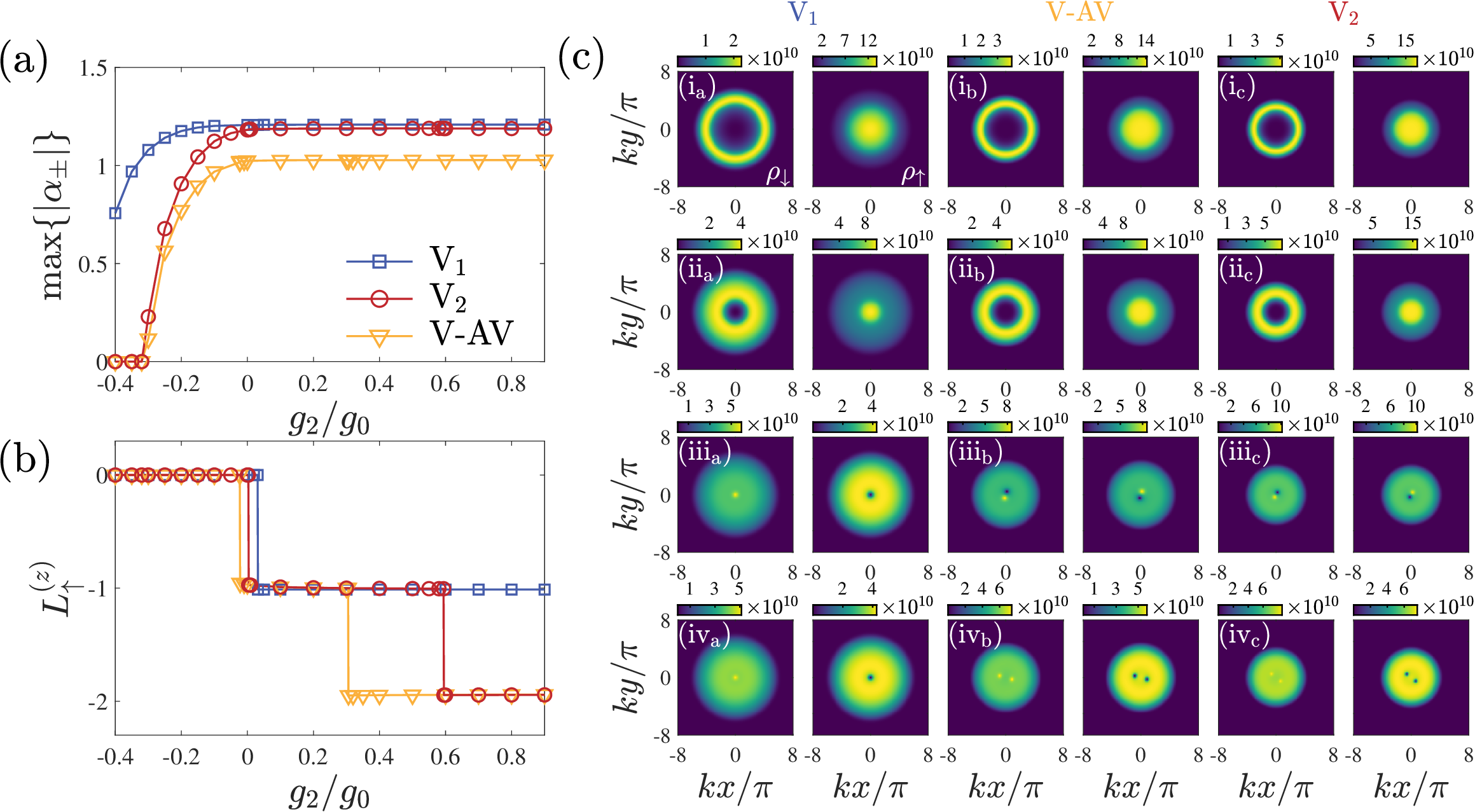}
	\caption{
		(Color online)
		Interaction-induced evolution of the chirality-polarized superradiant phases.
		(a) Maximum steady-state cavity amplitude $\max\{|\alpha_{\pm}|\}$ and (b) mean orbital angular momentum of the spin-up component
		$L_{\uparrow}^{(z)}$ as functions of the spin-dependent interaction $g_2$.
		(c) Representative density distributions of the spin-down (left) and spin-up (right) components for different interaction strengths.
		From left to right,
		$(\Omega_-,\Omega_+)=(3,2)E_L$,
		$(1,8)E_L$,
		and
		$(1,12)E_L$.
		The corresponding values of $g_2/g_0$ are
		($\mathrm{i_a}$) $-0.4$,
		($\mathrm{ii_a}$) $-0.2$,
		($\mathrm{iii_a}$) $0.2$,
		($\mathrm{iv_a}$) $0.4$;
		($\mathrm{i_b}$) $-0.2$,
		($\mathrm{ii_b}$) $-0.1$,
		($\mathrm{iii_b}$) $0.2$,
		($\mathrm{iv_b}$) $0.4$;
		and
		($\mathrm{i_c}$) $-0.2$,
		($\mathrm{ii_c}$) $-0.1$,
		($\mathrm{iii_c}$) $0.3$,
		($\mathrm{iv_c}$) $0.6$.
	}
	\label{Fig4}
\end{figure*}

\section{Stability of the superradiant phases}
\label{four}

Having established the emergence of chirality-polarized superradiance, we now investigate its robustness against two-body interactions, which are intrinsic to realistic Bose-Einstein condensates. The contact interaction takes the form~\cite{PhysRevLett.105.160403}
\begin{equation}
{\cal H}_{\rm int}
=
\frac12
\int d{\bf r}
\left(
g_0\hat n^2
+
g_2\hat S_z^2
\right),
\label{eq7}
\end{equation}
where $\hat n=\hat n_\uparrow+\hat n_\downarrow$ is the total atomic density and $\hat S_z=\hat n_\uparrow-\hat n_\downarrow$ is the spin density. The density interaction proportional to $g_0$ is SU(2)-spin-rotation invariant, whereas the spin-dependent interaction proportional to $g_2$ breaks the spin rotational symmetry. Assuming equal intraspecies interactions, $g_{\uparrow\uparrow}=g_{\downarrow\downarrow}$, the interaction parameters reduce to $g_0=g_{\uparrow\uparrow}+g_{\uparrow\downarrow}$ and $g_2=g_{\uparrow\uparrow}-g_{\uparrow\downarrow}$. Accordingly, the condensate is miscible for $g_2>0$ and immiscible for $g_2<0$.

Figure~\ref{Fig4}(a) and (b) summarize the interaction dependence of the cavity order parameter and the condensate orbital angular momenta for three representative superradiant phases. A striking feature is that the cavity order parameter remains nearly unchanged throughout the miscible regime ($g_2>0$), indicating that chirality-polarized superradiance is remarkably robust against moderate interaction-induced perturbations. 

By contrast, the cavity field is gradually suppressed as the system enters the immiscible regime ($g_2<0$). This suppression originates from the reduced spatial overlap between the spin components and the cavity-assisted Raman fields caused by interaction-driven phase separation, which weakens the collective Raman scattering responsible for superradiance. The influence of interactions becomes increasingly pronounced for higher-order vortex phases. As shown in Fig.~\ref{Fig4}(a), the doubly quantized vortex phase $\mathrm{V}_{2}$ loses stability at significantly weaker interactions than the singly quantized phase $\mathrm{V}_{1}$. This behavior reflects the larger centrifugal energy and stronger density depletion associated with higher OAM, rendering multiquantum vortices intrinsically more susceptible to interaction-driven instability. These results reveal a clear hierarchy of interaction effects: the collective cavity chirality remains robust, whereas the vortex topology is much more sensitive to interaction-induced reconstruction.

The corresponding density distributions are shown in Fig.~\ref{Fig4}(c). For the singly quantized phase $\mathrm{V}_{1}$, increasing $g_2$ drives the system toward the lower-energy vortex configuration while leaving the selected cavity chirality unchanged. In the immiscible regime, the condensate develops a pronounced annular density profile imposed by the LG Raman potential, allowing the singly quantized vortex to persist despite the reduced cavity occupation. A qualitatively different behavior is found for the higher-order vortex phases. For the doubly quantized phase $\mathrm{V}_{2}$, increasing $g_2$ first drives the system into the intermediate $\mathrm{V}$-$\mathrm{AV}$ phase and subsequently splits the doubly quantized vortex into two singly quantized vortices. The intermediate $\mathrm{V}$-$\mathrm{AV}$ phase therefore serves as the transition state connecting multiquantum and singly quantized vortex configurations. By contrast, in the immiscible regime the annular density profile generated by the LG Raman fields suppresses spontaneous vortex splitting and thereby stabilizes the doubly quantized vortex. The evolution of the $\mathrm{V}$-$\mathrm{AV}$ phase follows the same physical mechanism.

These results further distinguish the photonic and atomic responses to contact interactions. In the miscible regime considered, the dominant cavity chirality remains unchanged over a range of interaction strengths, whereas the condensate can undergo transitions between different vortex configurations. Thus, the atomic vortex topology can be modified without necessarily changing the selected cavity mode. Since the interaction strength can be continuously tuned via magnetic Feshbach resonances~\cite{RevModPhys.82.1225,PhysRevResearch.2.033152}, the present system provides a controllable platform for exploring interaction-induced topological transitions within chirality-polarized superradiant phases.

\begin{figure}[t]
	\includegraphics[width=1.0\linewidth]{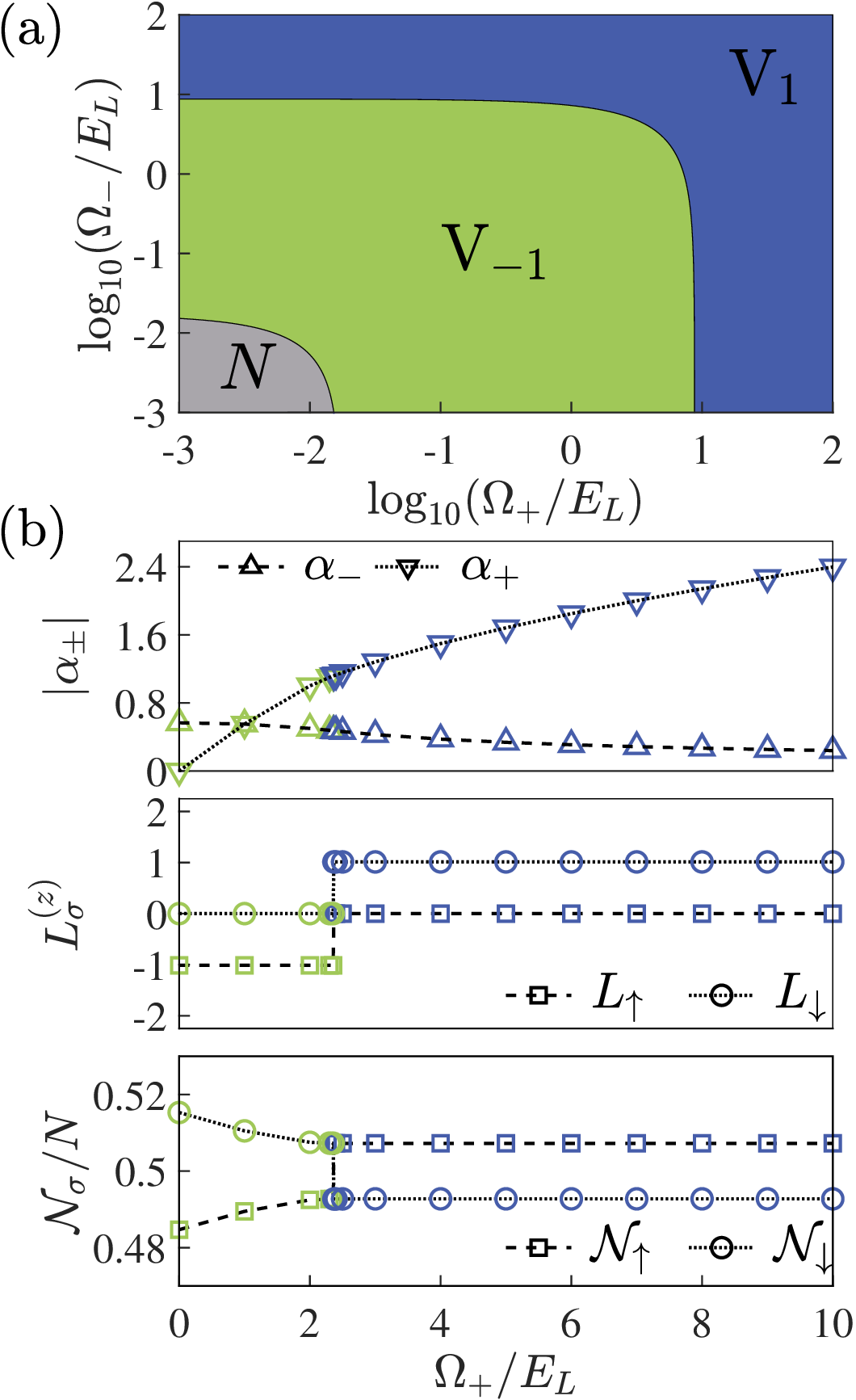}
	\caption{(Color online) (a) Ground-state phase diagram in the $\Omega_-$--$\Omega_+$ plane for equal OAM-transfer channels ($l_- = l_+ = 1$). (b) Steady-state cavity amplitudes $|\alpha_\pm|$ (upper panel), mean orbital angular momenta $L_\sigma^{(z)}$ (middle panel) and spin populations $\mathcal{N}_\sigma/N$ (lower panel) as functions of $\Omega_+$ for $\Omega_- = 1 E_L$.}\label{Fig5}
\end{figure}

\section{Role of quantized OAM transfer}
\label{five}

To elucidate the physical origin of chirality-polarized superradiance, we compare the unequal-OAM configuration with the equal-OAM case $l_-=l_+=1$. Figure~\ref{Fig5}(a) presents the corresponding ground-state phase diagram. Unlike the unequal-OAM case, both counterpropagating cavity modes become simultaneously populated throughout the superradiant regime. Because the two Raman channels have identical OAM dependence, they couple to the same orbital-angular-momentum sector without the OAM selectivity responsible for single-mode dominance in the unequal-OAM configuration. Consequently, the superradiant phase is characterized by the coexistence of both cavity chiralities with generally unequal occupations, corresponding to two-mode superradiance rather than chirality-polarized superradiance. When the remaining atomic and cavity parameters are symmetric under channel interchange, the equal-OAM configuration also admits an exchange symmetry under $(\Omega_-,\Omega_+)\leftrightarrow(\Omega_+,\Omega_-)$. Consistent with this symmetry, the superradiant threshold depends primarily on the combined Raman coupling strength $\Omega=\sqrt{\Omega_+^2+\Omega_-^2}$, rather than on the individual pump amplitudes.

Figure~\ref{Fig5}(b) further illustrates the evolution of the cavity fields along a representative cut. Unlike the unequal-OAM case shown in Fig.~\ref{Fig2}, both counterpropagating cavity modes remain macroscopically occupied throughout the superradiant regime. Their intracavity photon numbers evolve continuously with the pump strengths and are generally unequal because of the different Raman coupling strengths associated with the two driving fields. Since both Raman processes couple to the same angular-momentum sector of the condensate, they do not compete to select a unique cavity chirality. The corresponding superradiant transition is accompanied by the evolution of the condensate OAM from $L_{\uparrow}^{(z)}=-1$ to $L_{\downarrow}^{(z)}=1$, consistent with previous studies of ring-cavity SOAMC~\cite{PhysRevLett.120.123601}.

The comparison between Figs.~\ref{Fig2} and \ref{Fig5} demonstrates the role of quantized OAM transfer in controlling the cavity-mode occupations. For equal OAM transfer ($l_-=l_+$), both counterpropagating cavity modes remain populated in the superradiant regime considered. By contrast, unequal OAM transfer ($l_-\neq l_+$) couples the Raman channels to distinct atomic angular-momentum sectors, leading to competition between the corresponding atom--cavity scattering processes and favoring a single cavity chirality. The resulting mode selection provides an optical signature of the angular-momentum-dependent collective Raman response. By contrast, when $l_-\neq l_+$, the Raman channels become angular-momentum selective and couple to distinct orbital-angular-momentum sectors of the condensate. This symmetry breaking lifts the equivalence between the two cavity chiralities and locks the collective superradiant emission to a single running-wave cavity mode. These results establish quantized orbital-angular-momentum transfer as a new mechanism for controlling collective cavity superradiance through angular-momentum-selective Raman scattering, thereby extending the paradigm of ring-cavity superradiance beyond conventional dynamical SOC.

\section{Conclusions and outlook}
\label{six}

In summary, we have investigated collective light-matter interactions in a spin-$1/2$ Bose-Einstein condensate subjected to cavity-mediated SOAMC, which is generated by LG Raman beams in a two-mode optical ring cavity. We demonstrate that quantized OAM transfer controls collective superradiance through competing angular-momentum-selective Raman channels. The self-consistent feedback between atomic spin coherences and intracavity fields links the condensate orbital motion to the selection of cavity chirality. When the two Raman beams carry different orbital angular momenta, the cavity-assisted Raman processes couple to distinct angular-momentum sectors of the condensate,favoring one counterpropagating cavity mode and enabling chirality switching as the Raman pump strengths are varied. By contrast, identical orbital-angular-momentum transfer allows both cavity modes to coexist in the superradiant regime considered. These results identify quantized OAM transfer as a mechanism for controlling collective cavity emission beyond conventional Dicke-type superradiance~\cite{PhysRevLett.120.123601}, extending SOAMC from atomic orbital control to self-consistent cavity-chirality selection~\cite{PhysRevLett.122.110402}.

We further demonstrate that the interplay between cavity-assisted Raman coupling and contact interactions supports a rich family of topological superradiant phases with distinct vortex configurations and synthetic gauge-field textures. While the cavity chirality remains robust over a broad interaction range, the associated vortex topology undergoes interaction-driven reconstruction, revealing distinct responses of the photonic and atomic degrees of freedom. More broadly, our work provides a route to controlling collective light-matter states through quantized OAM transfer, with implications for chirality-selective superradiance~\cite{PRXQuantum.6.020101,Pucher2022}, interaction-driven topological dynamics~\cite{PhysRevResearch.2.033152}, synthetic gauge fields~\cite{doi:10.1126/science.aay3183, doi:10.1126/science.adf9621}, and chiral quantum optical phenomena~\cite{Lodahl2017,Zhang2025} in multimode cavity-QED systems.

A possible implementation employs a quasi-two-dimensional $^{87}$Rb condensate coupled to an optical ring cavity and driven by two external LG Raman pumps, requiring precise spatial mode matching and control of the Raman resonance conditions. The predicted chirality switching could be detected through mode-resolved cavity output, while the associated vortex structures could be characterized by spin-resolved imaging and interferometry. Comparing equal- and unequal-OAM configurations would provide a direct test of how quantized OAM transfer controls collective cavity-mode selection through self-consistent atom--cavity feedback.

\begin{acknowledgments}
This work was supported by the National Natural Science Foundation of China (Grant No. 12674338 and Grant No. 12374365), Quantum Science and Technology-National Science and Technology Major Project (Grant No. 2025ZD0300400) and Guangdong Provincial Quantum Science Strategic Initiative (Grant Nos. GDZX2604004 and GDZX2505001).
\end{acknowledgments}
 
\bibliography{reference.bib}
\end{document}